\documentclass[review,12pt]{elsarticle}

\usepackage{amssymb}

\usepackage{color} 
\usepackage{hyperref}

\journal{Physica A}

\begin{document}

\begin{frontmatter}



\title{Negative emotions boost users activity at BBC Forum}


\author[label1]{Anna Chmiel}
\author[label1]{Pawe{\l} Sobkowicz}
\author[label1]{Julian Sienkiewicz}
\author[label2]{Georgios Paltoglou}
\author[label2]{Kevan Buckley }
\author[label2]{Mike Thelwall}
\author[label1]{Janusz A. Ho{\l}yst \corref{cor1}}

\address[label1]{Faculty of Physics, Center of Excellence
for Complex Systems Research, Warsaw University of Technology,
Koszykowa 75, PL-00-662 Warsaw, Poland}
\address[label2]{School of Technology,
University of Wolverhampton, Wulfruna Street, Wolverhampton WV1 1LY, UK}

\cortext[cor1]{jholyst@if.pw.edu.pl}

\begin{abstract}
We present an empirical study of user activity in online BBC discussion forums, measured by the number of posts 
written by individual debaters and the average sentiment of these posts. Nearly $2.5$ million posts from over 18 
thousand users were investigated.  Scale free distributions were observed for activity in individual discussion 
threads as well as for overall activity.  The number of unique users in a thread normalized by the thread length 
decays with thread length,  suggesting that thread life is sustained by mutual discussions rather than by independent 
comments. Automatic sentiment analysis shows that most posts contain negative emotions and the most active users 
in individual threads express predominantly negative sentiments. It follows that the average emotion of longer threads 
is more negative and that threads can be sustained by negative comments. An agent based computer 
simulation model has been used to reproduce several essential characteristics of the analyzed system. The model 
stresses the role of discussions between users, especially emotionally laden quarrels between supporters of opposite 
opinions, and represents many observed statistics of the forum.
\end{abstract}


\end{frontmatter}


\section{Introduction}
The individual behavior of Internet users looking for information and browsing the web has been the subject of many 
papers, e.g.,  \cite{huberman98-1,barabasi05-1,dezso06-1,chmiel09-1,radicchi09-1,ratkiewicz10-1,goncalves08-1,
vazquez06-1}. Recently, social aspects of the web have become particularly important because of the 
growth of e-communities, i.e., groups interacting online using sites like Facebook and Twitter or 
blogs and forums \cite{kujawski07-1,mitrovic09-1}. As a result the Internet transfers not only information 
but also emotions. 
This has been the subject of several small scale studies \cite{rafaeli97-1,papacharissi04-1,derks08-1,kushin09-1}.
Rafaeli and Sudweeks \cite{rafaeli97-1} pointed out differences in emotional content due to 
different communication modes. They separated one-time, one-way comments from pairwise exchanges, 
further dividing the latter into reactive (where a message is a reaction to a preceding one) and interactive 
(where subsequent messages relate to the whole discussion). Analysing group discussions in Bitnet, 
Usenet and Compuserve they observed that reactive messages constitute over 50\% of the total studied 
number. Moreover, messages by the most active users were significantly more reactive than average. 
A similar analysis has been done in \cite{sobkowicz10-1}, where a human classification of a few thousand 
posts demonstrated that the growth of discussions in a popular Polish Internet forum was dependent on 
the degree of controversy of the subjects debated. 
The behavior of users of Internet communities has also been the topic of computer simulations, for example Ding and Liu \cite{ding10-3} analysed opinion evolution in a large bulletin board community. Depending on the topic of discussion, they observed varying states of agreement/disagreement between the users and posts. Simulations, based on game theory, have reproduced this behavior, depending on the ratio of opposing agents in the simulated group. 
Extending the approach beyond opinions, Schweitzer and Garcia \cite{schweitzer10-1} proposed an agent-based model of collective emotions in online communities, while Chmiel and Holyst \cite{ijmfc_chmiel} proposed a model of social network evolution driven by the exchange of emotional messages.  

Large scale studies of user behavior were facilitated by
the recent development of automatic sentiment detection \cite{prabowo09-1,pang08-2,thelwall10-1}. 
This has enabled investigations of emotions in massive e-communities \cite{cyberemotions,thelwall10-1,mitrovic10-2}. 
In \cite{thelwall10-1} women were shown to give and receive more positive comments than men in the social network
MySpace although there were no gender difference for negative comments. 
The latter point is in agreement with earlier observations by Evers et. al. \cite{evers05-1}, that in conditions of anonymity 
gender differences in the expression of anger are absent, contrary to face-to-face interactions.
The propagation of emotions in blogs was also studied in \cite{mitrovic10-2}, using the bipartite network methodology.    
 
In this work we show that negative emotions can drive Internet discussions. 
We focus on a statistical analysis of a large set of user comments and emotions detected by automated tools. 
Our goal is to understand the relationship between user activities and the emotions that users express in individual discussion threads and in all of their posts. 
We compare the observations with an agent-based simulation of a discussion board, which includes pairwise exchanges of comments as a significant factor determining activity statistics as well as emotional content. The model reproduces many of the characteristic features of the forum discussions, suggesting that their origins are akin to those derived from human analyses of smaller datasets \cite{rafaeli97-1,sobkowicz10-1}.
The paper is organized as follows. First we introduce the data source and analysis methods, followed by detailed descriptions of user activities and emotions. Next, a computer simulation model is described and its results compared with the observations. Lastly we discuss implications of the findings, especially the correlation between user activity and negative emotions. 

\section{Results}

\subsection{Dataset}
We base our analysis on the BBC \emph{Message Boards}, a public discussion Forum. The discussions cover a wide selection of topics, from politics to religion.
For this report, we focus on a subset of the available discussions that was found to have interesting emotional content: the \emph{Religion \& Ethics}\cite{BBCreligion} and World/UK News \cite{BBCworld} discussion Forums. The analysis spans 4 years, from the launch of the website (July/June 2005 respectively) until the beginning of the data extraction process in June 2009. Overall, $97,946$ discussion threads were analyzed,
comprising of $2,474,781$ individual posts made by $18,045$ users.

The \emph{emotional classifier} program that was used to analyze the emotional content of the discussions is based 
on a machine-learning (ML) approach. The algorithm functions in two phases: initially, during a \emph{training phase}, 
it is provided with a set of documents classified by humans for emotional content (positive, negative or objective) 
from which it \emph{learns} the characteristics of each category. Subsequently, during the \emph{application phase}, 
the algorithm applies the acquired sentiment classification knowledge to new, unseen documents. In our analysis, 
we trained a state-of-the-art hierarchical Language Model \cite{ounis08-1,peng03-1} on the Blogs06 collection 
\cite{ounis08-2} and applied the trained model to the extracted BBC discussions. Each post is therefore annotated 
with a single value $e=-1$, $0$ or $1$ to quantify its emotional content as negative, neutral or positive, respectively. 

\subsection{User activity}

In literature there are various observables introduced to characterize Internet user behavior. 
The analysis of inter-event time and waiting time distributions is very common
\cite{goncalves08-1,radicchi09-1,barabasi05-1,vazquez06-1,masuda09-1}, and these can be described by power-law 
relationships. Barab\'{a}si \cite{barabasi05-1} suggested that the bursty nature of various human activities in cyberspace 
(e-mail, web-browsing) follows from decision-based queuing processes. Radicchi \cite{radicchi09-1} found that 
the distribution of inter-event times for a user is strongly dependent on the number of operations executed by that user.

\begin{figure*}[!ht]
\centerline{\begin{tabular}{c c c }
\includegraphics[width=2in]{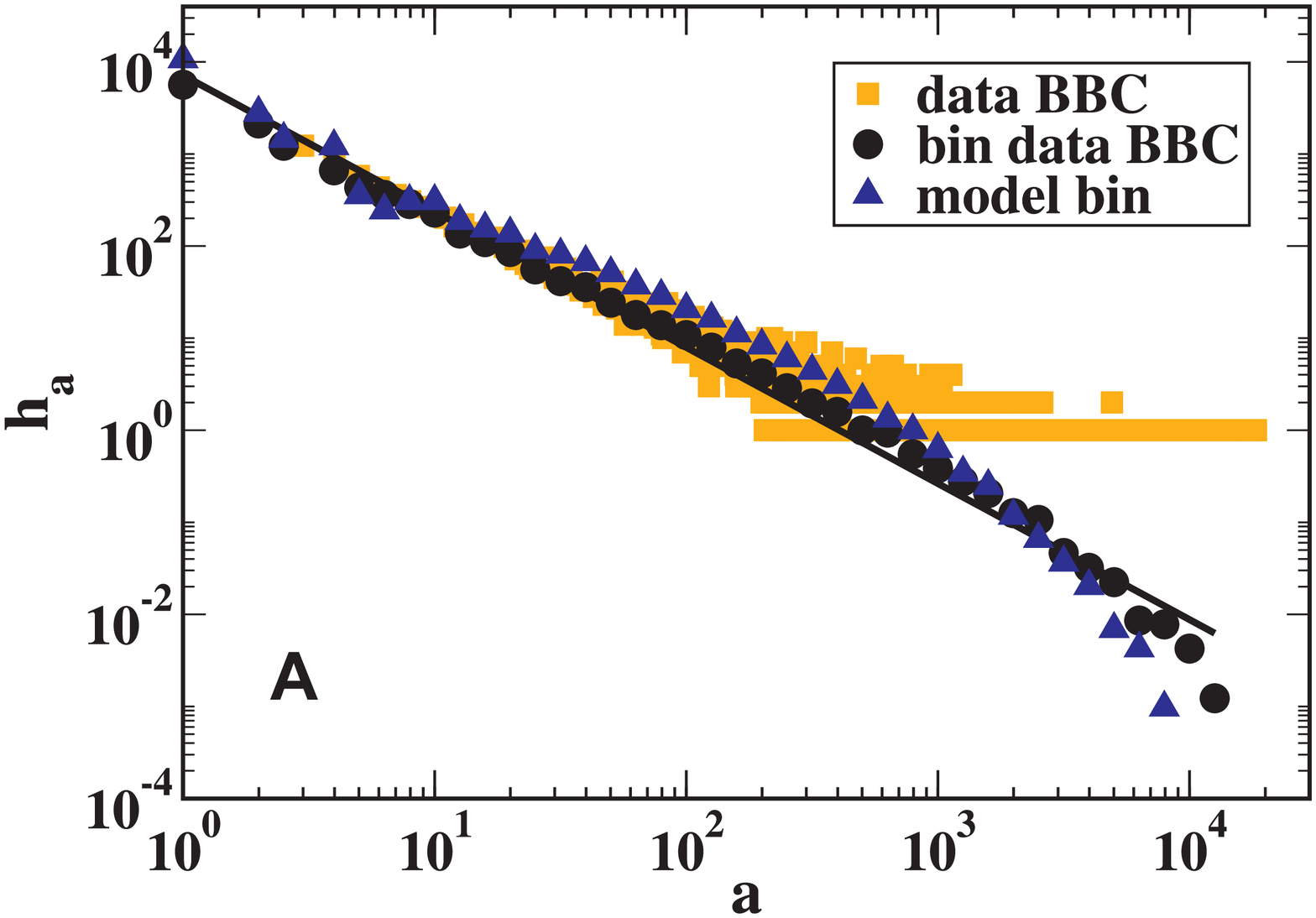}&\includegraphics[width=2in]{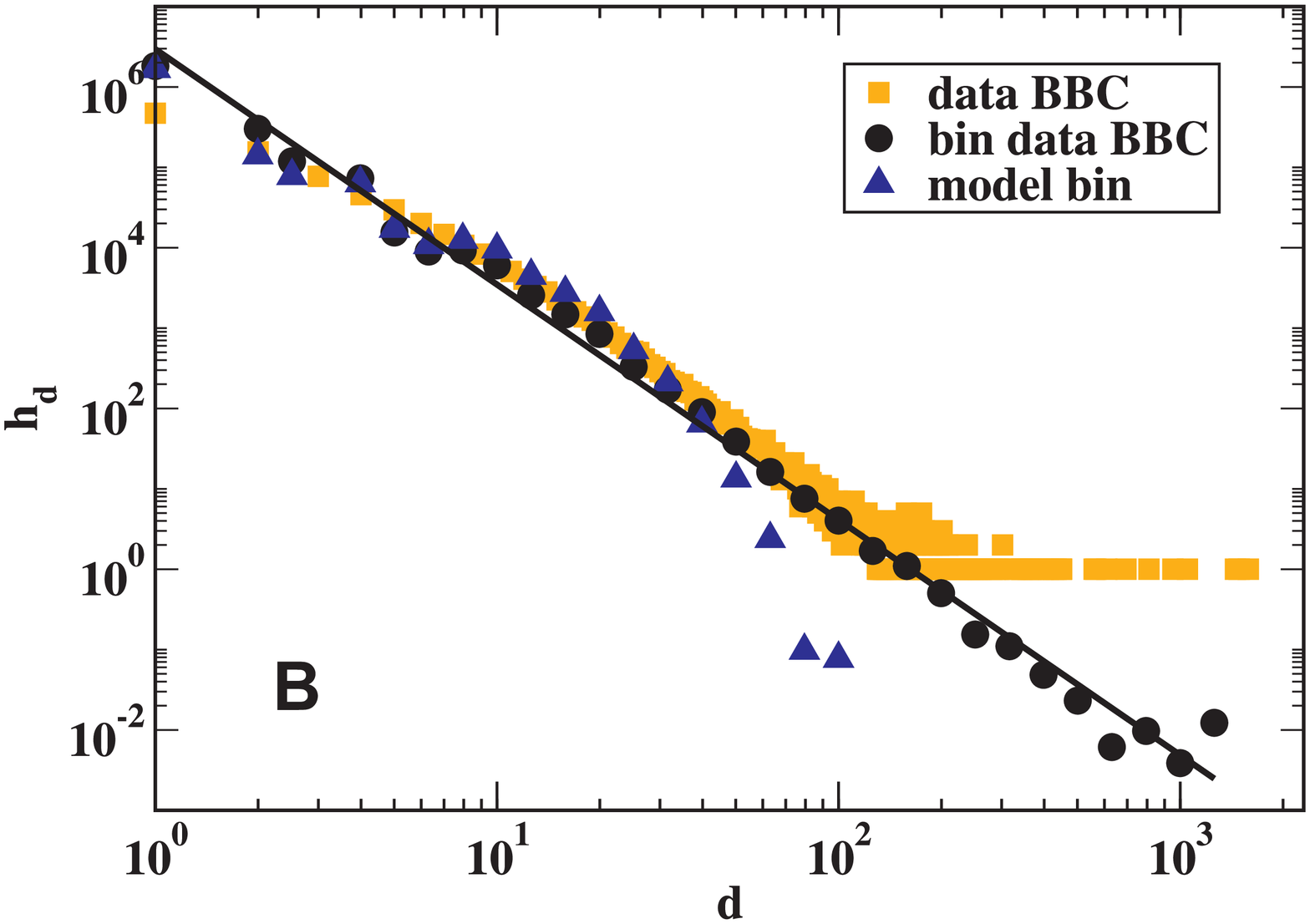}  & \includegraphics[width=2in]{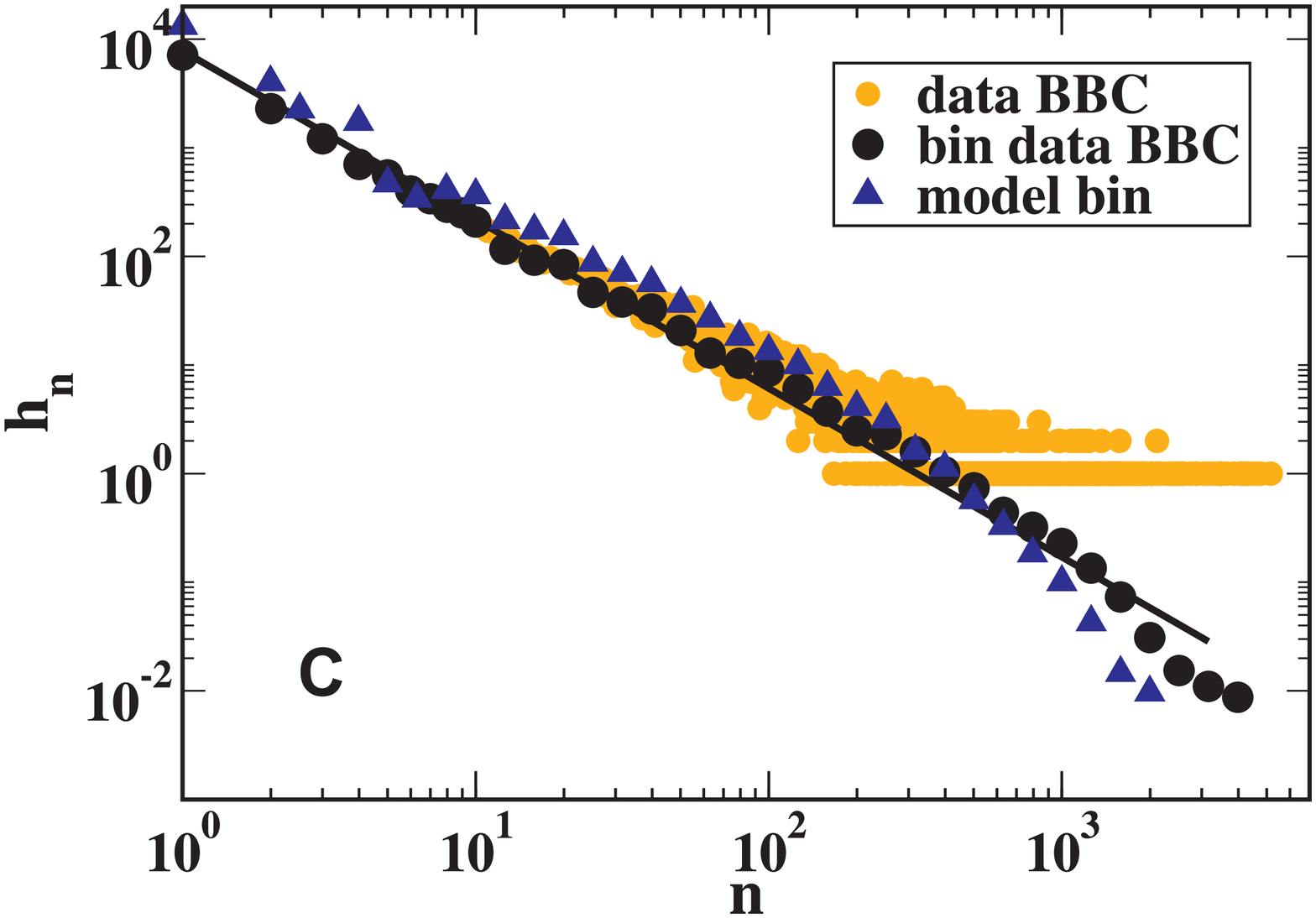} 
\end{tabular}}
   \caption{ (Color on-line) Histograms of user activity and diversity of threads: (A) Histograms of user total activity $a$ (orange squares), binned data (black circles), the binned result of the numerical simulation of the agent model (blue triangles) (B) Histogram of user activity in thread $d$ (C) Histogram of the diversity of threads $n$. Lines are fits to the data and they follow $h_a \sim a^{-\beta}$, $h_n \sim n^{-\tau}$, $h_d \sim d^{-\alpha}$ with $\beta=1.4$, $\tau=1.5$, $\alpha=2.9$. }
   \label{1}
\end{figure*}

Here we consider user activity $a_i$ defined to be the total number of posts written by user $i$ in 
all discussion threads during the observation period. For simplicity, this quantity will also be referred to as $a$. 
The maximum observed activity in the dataset is $a_{max}=18274$, i.e.,  one user authored more than eighteen 
thousand messages, while the average activity is $\langle a \rangle = 137$, and the median is $m_a=3$. 
The number of occurrences of $a$ is illustrated in Fig. \ref{1}(A) (and it is well fitted by the power-law 
relationship $h_a \sim a^{-\beta}$, where $\beta = 1.4$. The relatively small value of the exponent $\beta$ 
suggests a high number of very active users of this Forum. All figures showing histograms of user activity have the same layout:
the real data and the numerical simulations of the model (see section \ref{SecSimul}).

Since all discussions in the Forum are split into separate threads $j$, we define $d_{i}^j$ (or $d$ for short) to be the 
local activity of user $i$ in thread $j$ measured by the number of posts that the user submitted to the discussion.  
Although both its maximum and average value (respectively $d_{max}=1582$ and $\langle d \rangle = 2.84$) are lower than in 
the case of $a$, the number of occurrences of $d$, shown in Fig. \ref{1}(B), still follows a similar relationship $h_d \sim
 d^{-\alpha}$ with exponent $\alpha=2.9$, which is double that of $a$. 

Taking into account the above described quantities: (1) How many threads are in the area of interest of a user? 
(2) How does the user spread her activity among different discussions? To answer these issues, consider the number of 
different discussions $n_i$ in which the user $i$ takes part. The results in Fig. \ref{1}(B) show the number of occurrences of 
$n_i$. Again we find power-law scaling $h_n \sim n_i^{-\tau}$ with $\tau=1.5$. The results reveal diversity in human 
habits: the overwhelming majority of users join just one discussion and usually post only one comment in it. However, 
there is also a significant number of those that write often and express themselves in several discussions.

\begin{figure*}[!ht]
\centerline{\begin{tabular}{c c c }
\includegraphics[width=2in]{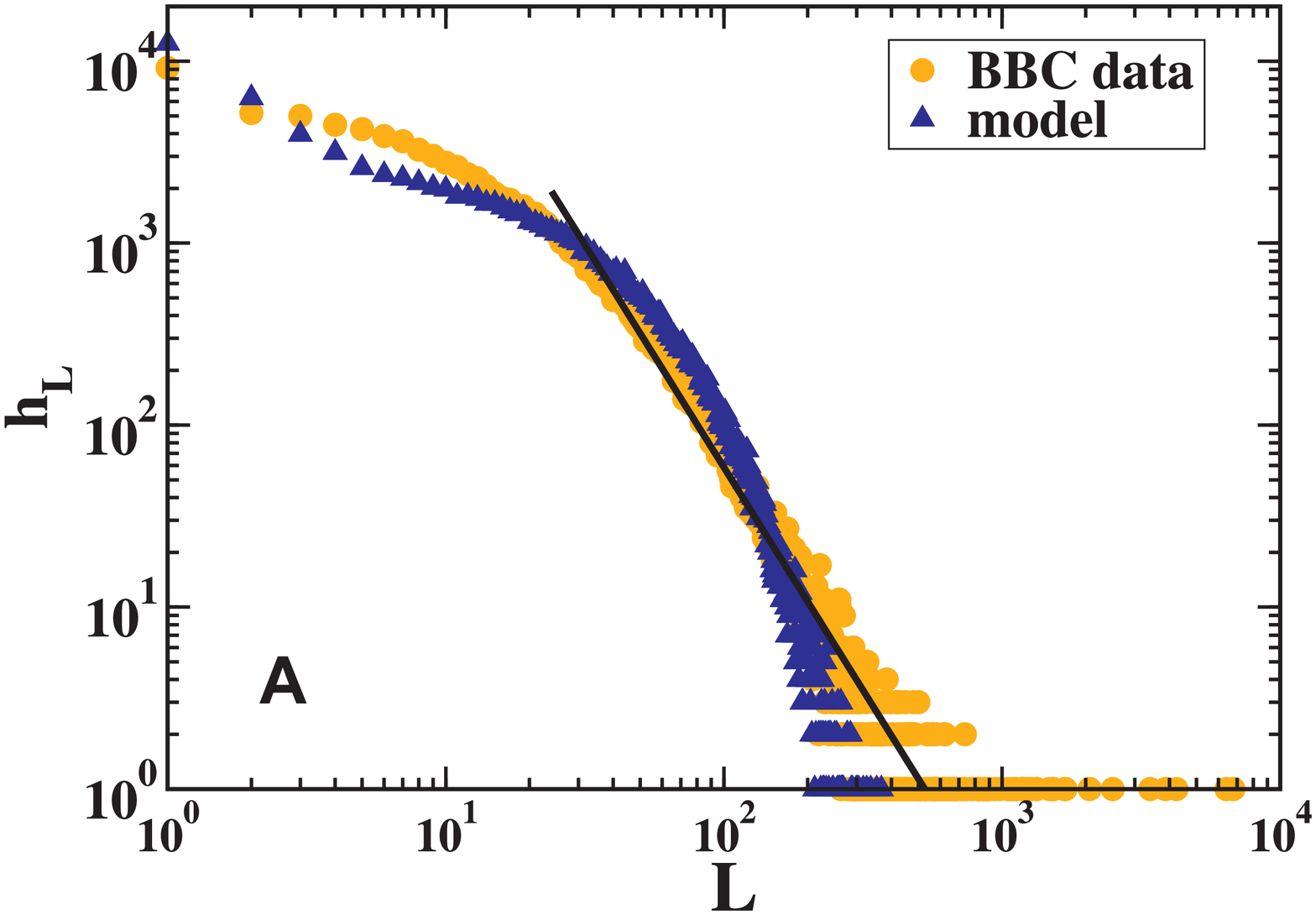}  &\includegraphics[width=2in]{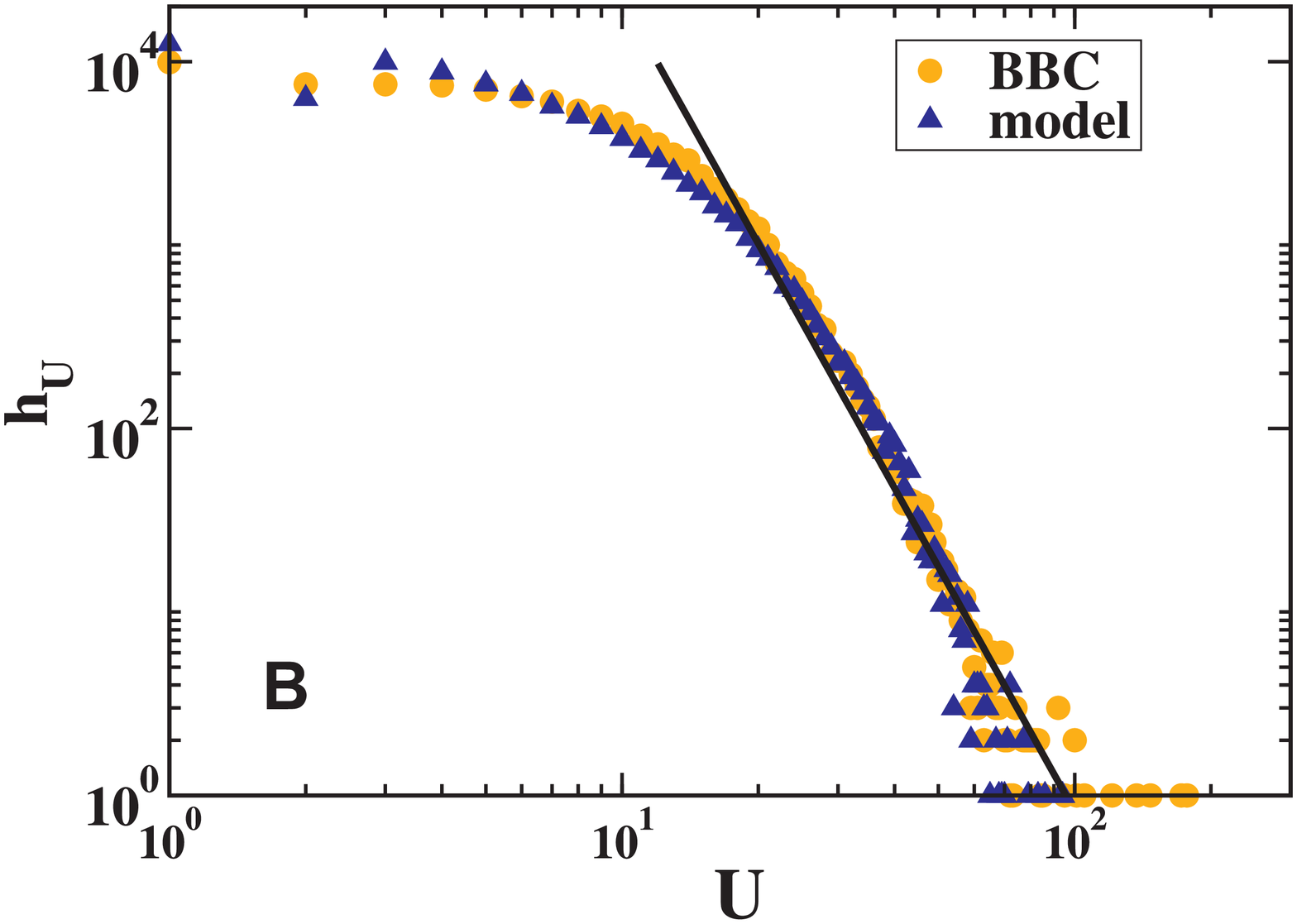} &  \includegraphics[width=2in]{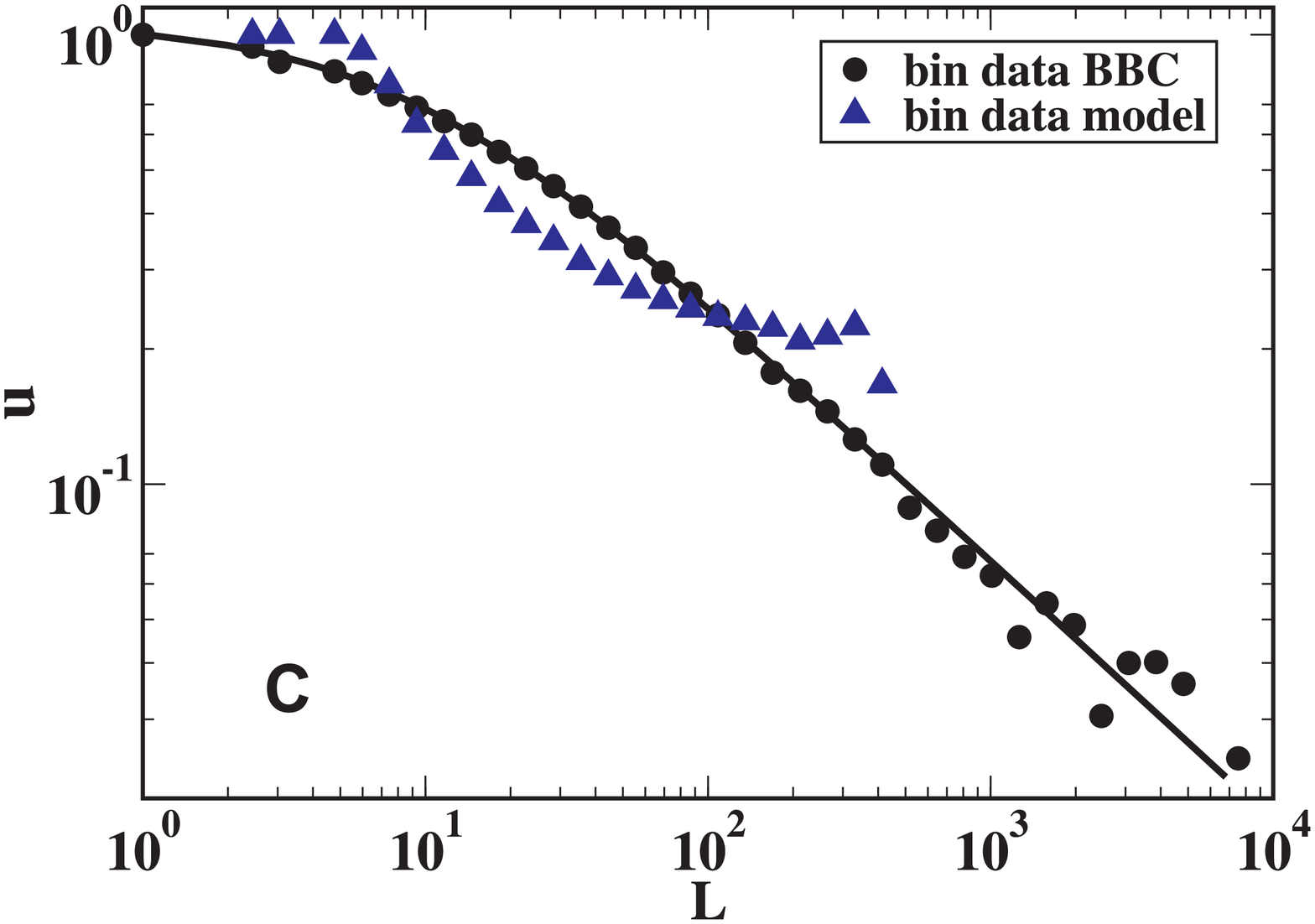} 
\end{tabular}}
   \caption{(Color on-line) (A) Histogram of threads with length $L$. The BBC data are represented by orange circles, the result of the numerical simulation of the agent model is shown by blue triangles; the black line corresponds to $h_L \sim L^{-\eta}$ with $\eta=2.5$ (B) Histogram of the number of unique users $U$ making a comment in the thread. Lines correspond to relation $h_U \sim U^{-\eta}$ with $\eta=4.9$ (C) The normalized number of unique users $u$ making a comment in a thread of length $L$, black squares - data binned logarithmically, blue triangles - model. The black line corresponds to $u=A(L+b)^{-\delta}$ with fitted parameters $\delta=0.58$, $A=3.72$ and $b=8.6$.} 
   \label{2}
\end{figure*}

Although statistical behavior of users shows a strong tendency to be scale-invariant, this is not so clear for the thread statistics shown in Fig.~\ref{2}(A). Here, we consider the thread length $L$ and the number of unique users $U$ posting at least one comment in the thread.  Histograms of both $h_L$ and $h_U$ display power-law tails for $U,L > 20$. This is most prominent in the case of $h_U$, which is also characterized by a rather large exponent  $\eta=4.9$.

To understand the impact exerted by the most frequent users on the length of a thread consider the dependence 
between the normalized number of unique users in a single thread defined by $u=U/L$ and thread length 
(Figure \ref{2}(C)).   For short threads ($L$ between 1 and 10) $u$ is about $0.6-1$ while for threads 
larger than $400$ comments it drops below $0.1$. A good fit is $u(L) =A(L+b)^{-0.58}$ 
(the black line in Fig. \ref{2}(B)) thus the number of unique users grows more slowly than linearly
 with thread lengths. This suggests that mutual discussions between specific users 
rather than a large number of independent comments submitted by many users sustain thread life.

\subsection{User emotions}

As mentioned in the Introduction, the recent progress in automatic sentiment analysis gives the ability to 
quantify the emotional content of large scale textual data. This has already led to observations of 
emotionally-linked communities in blogs \cite{mitrovic10-2} and to tracing shifts in public opinion 
\cite{gonzalez-bailon10-2}. Other, indirect methods have also revealed the emergence of phenomena like 
the existence of the `hate networks' in political discussions \cite{sobkowicz10-1} and emotional connections 
within communities in massive multi-player online games \cite{szell10-1}. Kappas \emph{et al.} 
\cite{kappas10-1} have demonstrated that BBC Forum posts elicit physiological reactions consistent with the apparent emotional content of the messages in people that read them when participating in psychological tests.

\begin{figure*}[!ht]
\centerline{\begin{tabular}{c c  }
\includegraphics[width=3in]{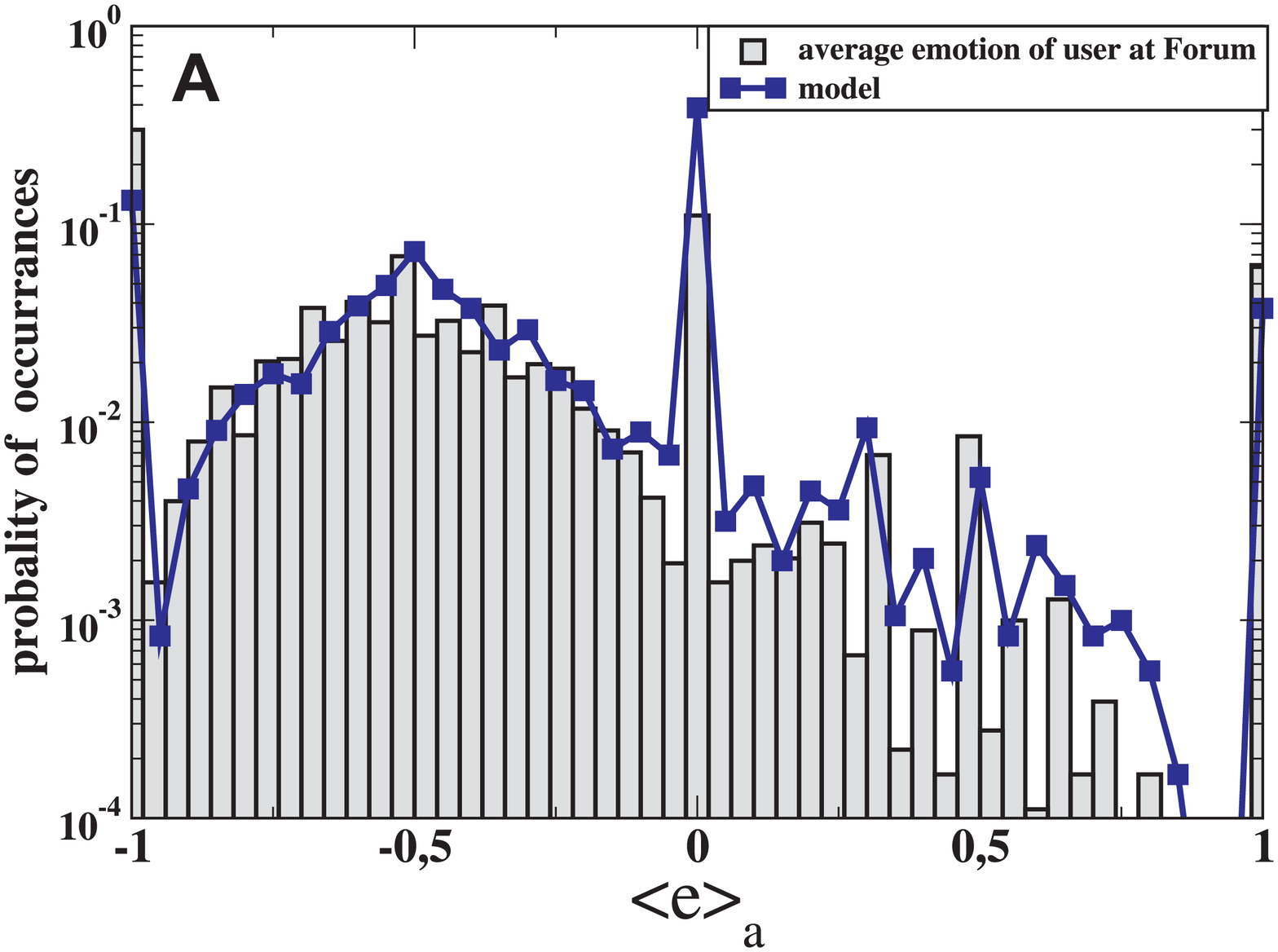} & \includegraphics[width=3in]{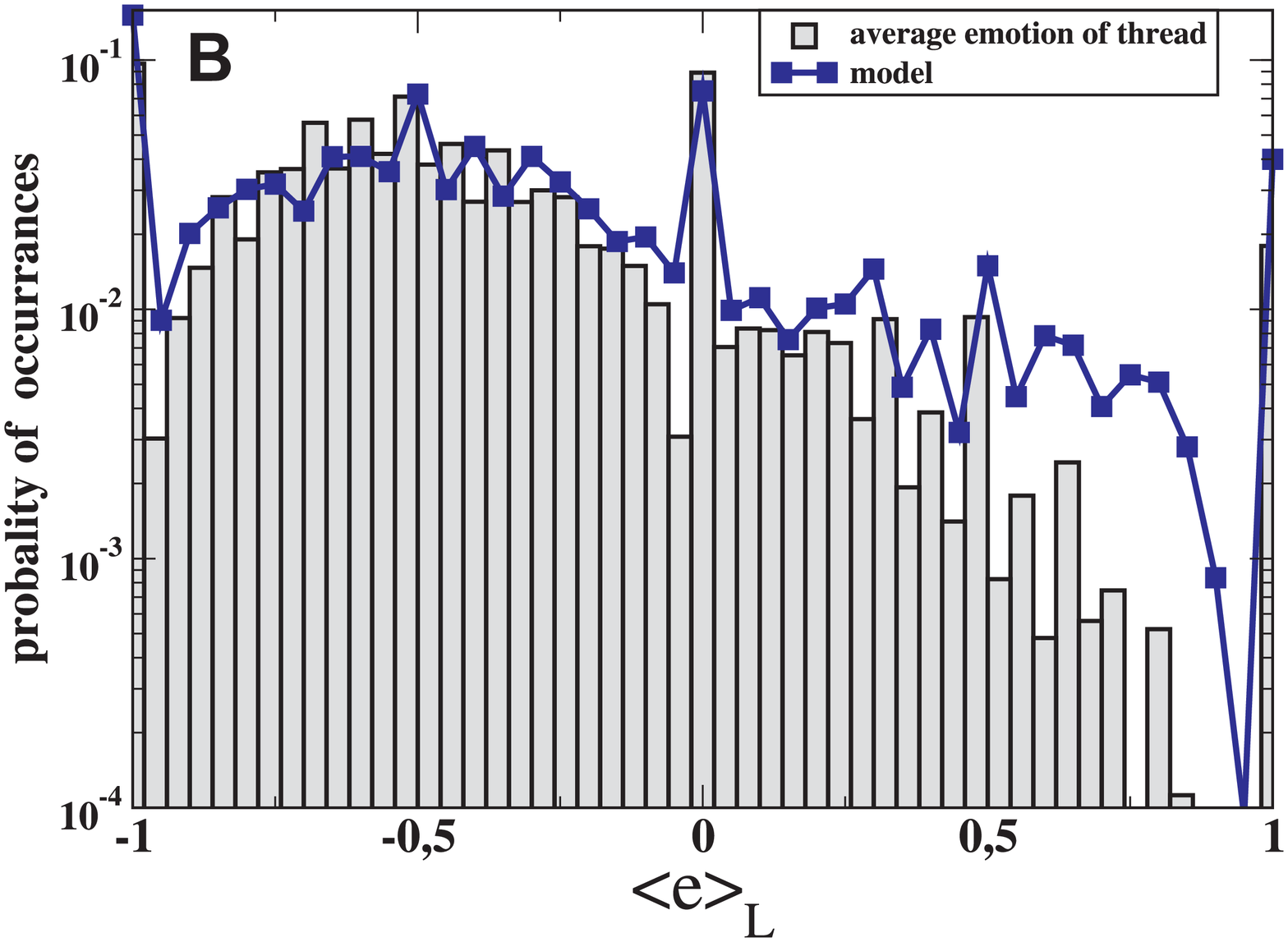}
\end{tabular}}
\caption{(Color on-line)Distribution of emotion. (A) Probability distribution of users' global average emotions $\langle e \rangle_a$ and (B) global emotion of threads $\langle e \rangle_L$. BBC data is represented by gray bars, the result of numerical simulation of the agent model by the blue squares; the line is only to guide the eyes.}
\label{3}
\end{figure*}

The following quantities describe the emotions of individual debaters and discussions threads. 
The average (global) emotion of a user $\langle e \rangle_a$ is the sum of all emotions $e$ in posts 
written by the user $i$ divided by her activity $a_i$. The average emotion of a thread  
$\langle e \rangle_L$ is the sum of all emotions in the thread $j$ divided by its length $L_j$. 
The third value $\langle e\rangle_d$ is the average emotional expression of the user $i$ in the thread $j$. 
The main features of the distribution $p(\langle e\rangle_a)$, shown in Fig. \ref{3}(A),  
are peaks for $\langle e\rangle_a=-1, 0, 1$ which are a straightforward effect of the large number of 
users with $a=1$ and threads with $L=1$ (see Fig. \ref{1}(A) and Fig. \ref{2}(A)). 
The local maximum around $\langle e\rangle_a=-0.5$ is a specific attribute of the BBC Forum because it 
possesses a strong bias toward negative emotions, with an average value of $\langle e \rangle = -0.44$. 
Fig. \ref{3}(B) contains the distribution of $p(\langle e \rangle_L)$. A similar shape is present 
for $p(\langle e \rangle_d)$. 

\begin{figure*}[!ht]
\centerline{\begin{tabular}{c c }
\includegraphics[width=3in]{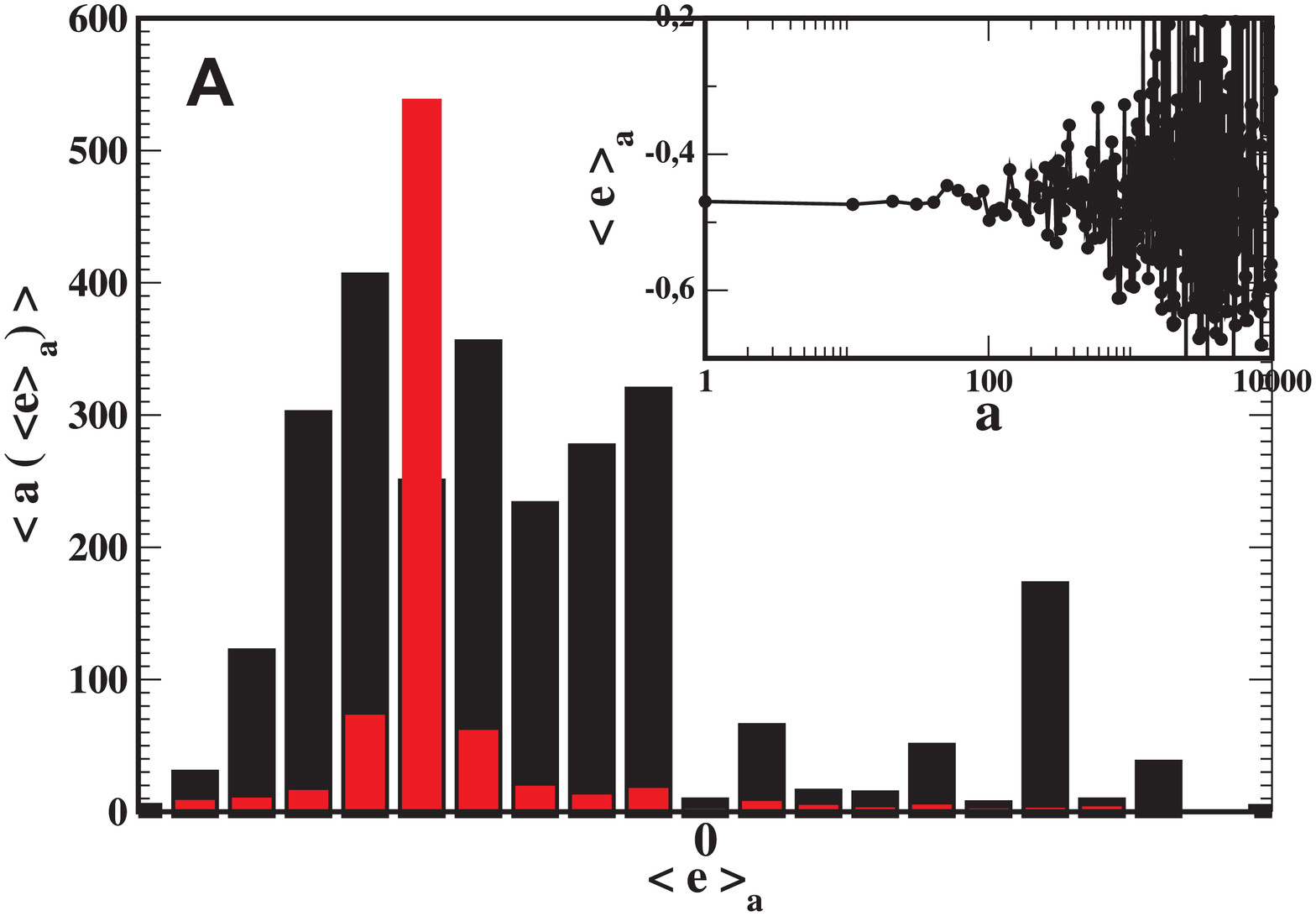} & \includegraphics[width=3in]{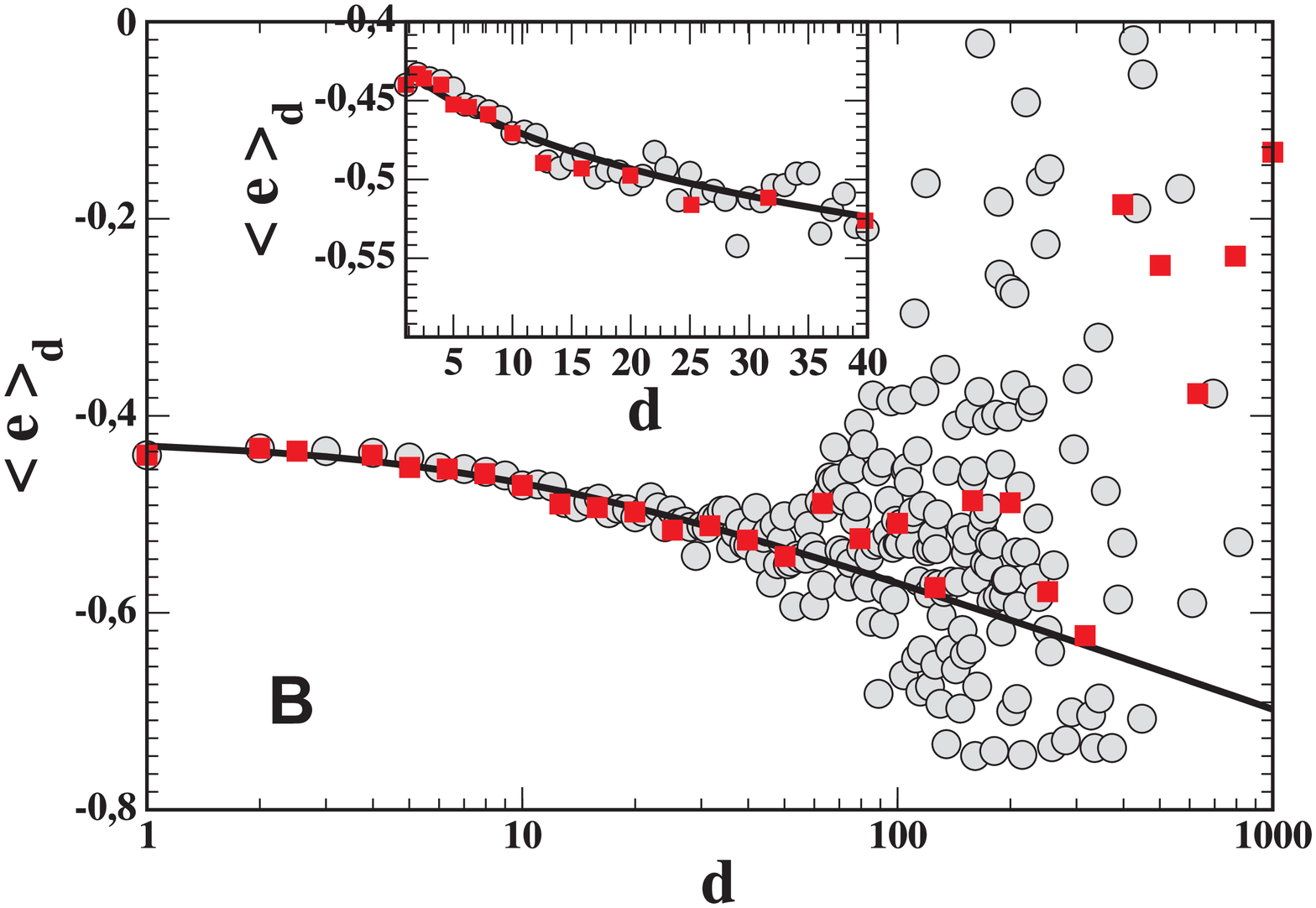}
\end{tabular}}
\caption{(Color on-line) (A) Average activity of a user as a function of the user's emotions. Users' global activity $\langle a \left( \langle e \rangle_a \right)\rangle$  versus their global average emotion $\langle e \rangle_a$: black bars - empirical data, red bars - shuffled data. Inset: Users' global average emotion $\langle e \rangle_a$ versus users' global activity $a$. (B) Relationship between users' average emotion in a thread $\langle e \rangle_d$ and users' activity in the thread $d$. Grey circles are original data, red squares are binned data and the black curve corresponds to equation $\langle e \rangle_d=A_1+B_1 \ln(d+b)$ with $A_1=-0.31$, $B_1=-0.054$ and $b=8.6$.  The inset is focused on local activity in threads with less than 40 messages from a single user in  one thread.}
\label{4}
\end{figure*}

So far we have treated user activities and emotions as mutually independent variables but consider now 
the relationship  between them.  Fig. \ref{4}(A) (inset) graphs users' global average 
emotions $\langle e \rangle_a$ versus their global activity $a$. Neglecting fluctuations for large values 
of $a$ caused by small numbers of very active users, there is a constant mean emotion that is around the 
Forum's average value $\langle e \rangle$. Hence, on average, the user activity level $a$ does not influence 
her emotions $e$. In the main panel of Fig. \ref{4}(A) the reversed relationship is plotted, i.e., the 
distribution of average global activity versus users' average emotions (black  bars). 
For comparison we present shuffled data (red bars) where the emotional values of posts were randomly 
interchanged between users. Whereas the second distribution follows a Gaussian-like function, 
the original set is characterized by a broad maximum stretching across almost all of the negative 
part of the plot and some minor fluctuations in the positive part. 
This implies that there is a dependence between user activity and emotions even at the global level.

Users can take part in many threads, thus their local and global activities as well as corresponding 
local emotions can be very different.  But how are users' emotions $\langle e_i^j\rangle $ expressed in 
a thread connected to the activity level  $d_i^j$ in it? Fig. \ref{4}(B) shows the average emotions of a 
user in a thread as a function of the user's local activity. In this case, an increase in activity in a 
particular thread leads to more negative average emotions in the thread. Recall that there was no relationship 
between a user's global activity and her emotions, as shown in the inset of Fig. \ref{4}(B)). 
For longer discussions there is a more homogeneous group of users (see Fig.\ref{2}(C)),
thus on average one user writes a larger number of posts $\langle d
\rangle(L)=\frac{1}{u(L)}$ in such a thread as compared to shorter ones.
As shown in Fig. \ref{4}(B), the average emotions for users that are locally more active decreases. These two effects document that the longer threads possess, 
on average, more negative emotions. In fact in Fig. \ref{ke} there is a logarithmic decay in mean 
thread emotions $\langle e\rangle _L $ as a function of thread length $L$. 

\subsection{Computer model and simulations}
\label{SecSimul}

To analyze our data we have developed an extension of simple computer model of
 the community and discussion process. Such simulations, using agent-based computer models, 
have been previously proposed in \cite{schweitzer10-1,ding10-3,sobkowicz10-1,czap}. 
We use a modification of the model introduced in \cite{sobkowicz10-1}, which may be described as follows. 
The simulated users (agents) belong to three categories: those with deeply set, 
controversial opinions (denoted as A and B), and neutral agents, denoted as N. 
Comments always reflect opinion of the user, so we would use the same categories to describe users and posts.
As the simulation makes no attempt to map real topics of BBC discussions, we have no knowledge regarding 
the relative numbers of disagreeing factions A and B. For this reason we have assumed that the 
size of both opinionated groups is symmetrical. 
Independent of the user community mix, the topics of discussion (called source messages) 
may also favor A, B or a neutral viewpoint. 
 
 It should be noted that the comment opinion (A/B/N) is separate from the comment emotional content.
The latter is determined by the comparison between categories of its author and the target 
of the post (which may be the source message of the thread or another post). Adding neutral agents allows more 
flexibility in treating emotions than in the original model in Ref.~\cite{sobkowicz10-1}. 
Neutral agents' posts are always emotionally neutral.
If the author of the comment and the target
belong to the same non-neutral category (A-A, B-B) then the emotion expressed in the comment is positive. 
For A-B or B-A combinations emotions are negative. For A or B category authors commenting on neutral posts 
there is certain probability $x_N$ that the comment would be emotionally negative, and in remaining cases 
it would be neutral. In contrast to Ref.\cite{sobkowicz10-1} we used only one class of agent activity.

We have assumed a population of 25000 agents reading the forum. For each thread we randomly select agents who may 
participate in the discussion. Each such agent `reads' one of the messages within a thread (called the target). 
The target may be the source of 
the thread (usually news item from BBC) or an earlier post by another user. 
The probability of the agents to read the source message, $p_s$, 
is one of the control parameters of the simulation. 
With probability $(1-p_s)$ the target is another earlier post, where we assume preferential attachment rules to calculate the probability of a reader choosing a specific comment as the target. 
Specifically, the chance of reading a post is proportional to its total degree 
(the outdegree of a post is always 1, but the indegree may be quite high).
After reading the target post, the agent then decides whether or not to comment on it.
 
The probability of posting a comment is given by $p_c f(r,t)$, where a universal `comment activity' ratio $p_c$, 
the same for all agents, is modified by a factor $f(r,t)$, depending on the reader $r$
category and that of the target message $t$. This factor reflects a greater probability of 
getting aroused by contrary views and becoming motivated to post a comment. 
Thus for ($r=$A, $t=$B) and ($r=$B, $t=$A) pairs we have $f(r,t)=1$, while for other combinations $f(r,t)=f^*<1$, where $f^*$ is an adjustable parameter.

After the agent has commented on a post other than the source we enter into a `quarrel' subroutine. Here, the author 
of the target post is `given a chance' to respond, with probability determined by $p_r f(r,t)$, where $p_r$ is 
an independent parameter from $p_c$, but $f(r,t)$ is the same as in the main routine. 
If the response is placed, the roles of the two agents are reversed, and a chance for counter-response is evaluated. 
This subroutine continues until one of the agents `decides' to quit. Values of $p_c$ and $p_r$ 
determine the relative importance of quarrels within the thread.

\begin{figure*}
\vskip 0.9cm
\centerline{\includegraphics[width=6in]{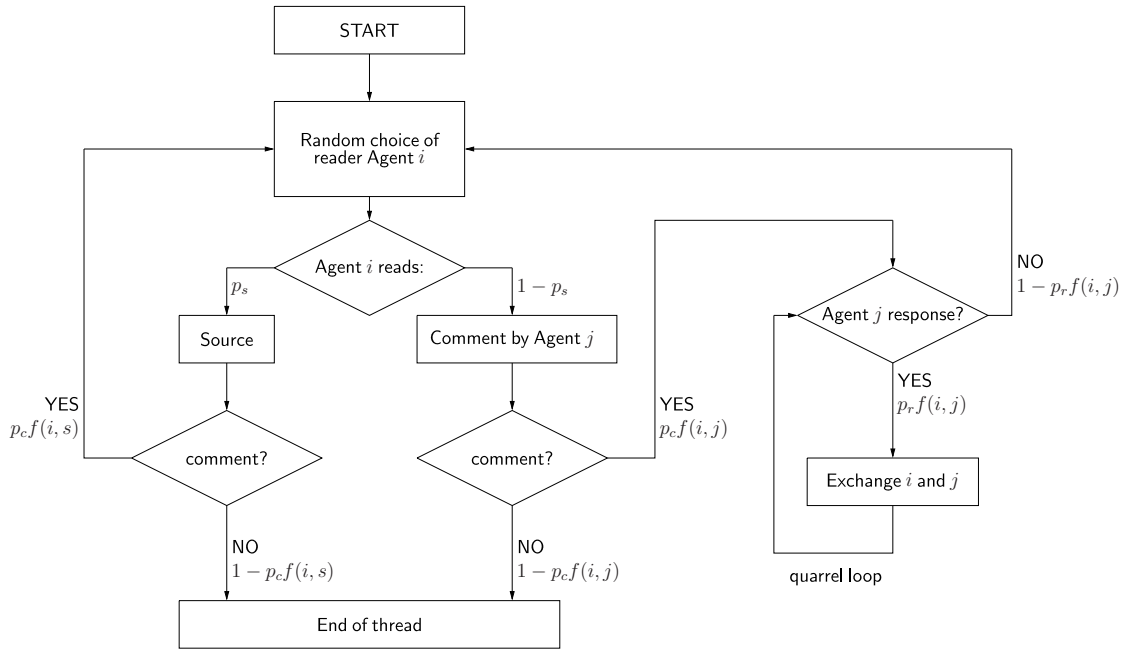}}
\caption{ Schematic diagram of the simulation process within a single thread. The process was repeated 110 000 times. Factor $f(reader,target)$ depends on combinations of opinions of the reader and the targeted post. For combinations where reader and target 
posts have different, non-neutral opinions $f(A,B)=1$ and $f(B,A)=1$, 
for other situations $f(reader,target)=f^*<1$. Using parameters  $p_s=0.5$, $p_c=0.93$, $p_r=0.89$ and 
$f^*=0.86$  resulted in $\sim 97000$ threads with at least one comment.}
\label{diagram}
\end{figure*}

This simple simulation program returns then to the main routine of agents posting comments, 
until the currently selected agent decides not to post. 
The whole process is then repeated for a specified number of threads $N_{th}$. Fig.~\ref{diagram} 
illustrates the flow of the simulation program for a single thread.

Our model does not include many features present in Internet discussions.  Emotions are automatically 
determined by the category of agents/posts, rather than resulting from the emotional content of previous messages. 
In real discussions it is commonly the use of offensive language that spurs replies in kind, not the opinion expressed in the post.
Also, situations such as a single user posting more than one post within a thread (due to, for example, 
lack of space to express his/her viewpoint or repetition of the same message several times) are not considered. 
This leads to deviations of the model statistics from observations for threads with only a few comments. 
Despite the model simplicity it gives results quite similar to the ones derived from our analysis of the BBC forum. 

\begin{figure*}
\vskip 0.9cm
\centerline{\includegraphics[width=4in]{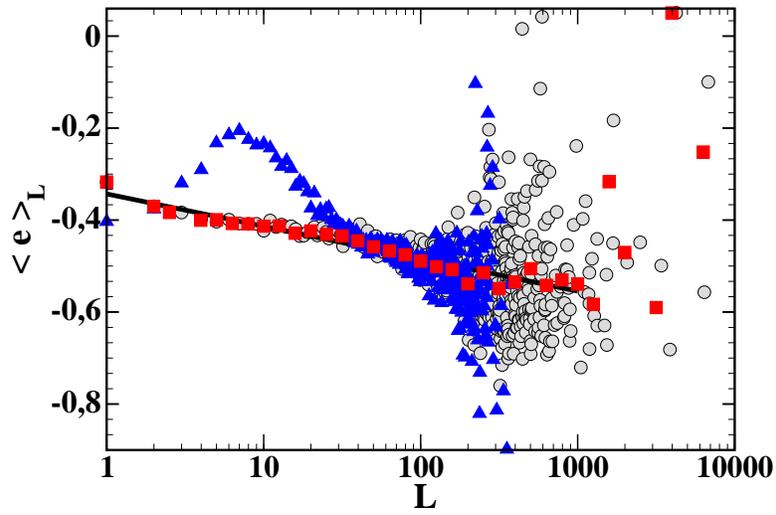}}
\caption{ (Color on-line) Average emotion in discussions with fixed thread length $L$ (gray squares): red circles are binned data, the black line corresponds to the relationship $\langle e \rangle_L = A' + B' \ln(L)$ with fitted parameters  $A'=-0.34$ $B'=-0.03$, blue triangles represent the model results.}
\label{ke}
\end{figure*}

Figures \ref{1}--\ref{3} and \ref{ke} compare the results of simulations obtained 
for $N_{th}=110000$, $p_s=0.5$, $p_c=0.93$, $p_r=0.89$, 
$f^*=0.86$ and $x_N=0.91$ with statistics of the BBC forum. 
This choice of parameters resulted in simulations with about 2.5 million posts, $\sim$970000 active 
threads (i.e. thread topics to which at least one agent posted a comment) and $\sim$19000 active agents (i.e. agents that posted  at least one comment) 
-- these values vary slightly between simulation runs, but correspond well to the observed data.
The results of simulations are obtained using relative sizes of communities supporting A, B and 
neutral viewpoints of 
A/B/N$=$32\%/32\%/36\%. For the topics of the threads (source messages) the 
fit resulted in a ratio of (A/B/N)$_{source}=$25\%/25\%/50\%.
 
About 70\% of the posts in the simulation belonged to quarrels. The choice of $f^*$ and $x_N$ was determined by the decision 
to obtain the same mean value of emotion $\left<e\right>$ as that measured in the BBC forum. 
The A/B/N ratios for users and source messages were chosen to reproduce the ratios of emotionally positive, 
neutral and negative messages.
The simulation resulted in values of 20\%, 16\% and 64\%, in good agreement with data from BBC forum: 
19\%, 16\% and 65\%, respectively. The distributions of average user and thread emotions (Fig.~\ref{3}) 
follow closely the observed statistics.
Other statistical characteristics of the discussions, such as user activity, thread lengths, 
unique authors, and thread diversity, are also rather close to the observed distributions. 

The differences between simulation and reality are visible mainly for short threads.  
 
For example, the distribution of user activity within a thread, while being in the same range as the data from 
the BBC forum, does not show power law behavior, with smaller counts for low activity values (Fig.~\ref{1}(B)).
Another difference is seen in the relationship between thread length and average emotion (Fig.~\ref{ke}), 
where the simulation deviates from observations for low values of $L<30$, but is close to the observations 
for longer threads. A broader analysis of sensitivity of numerical results on model parameters is presented in Appendix.

\section{Discussion}

Most of the previous studies documenting the role of emotions in computer mediated communications have been based on small 
scale samples of data or experiments. Our study is based on a very large, multi-year dataset, documenting 
the behavior of many users in online BBC fora.
Moreover, the topics covered by the analyzed fora are of significant social importance. 

Thus understanding why negative emotions dominate discussions and participating users may have important 
consequences in democratic processes. Because an increasingly large part of the information and opinions on which 
we base our decisions comes from the Internet, knowledge of mechanisms that may increase or decrease emotional 
content may help efforts to minimize social conflict and achieve consensus \cite{khatib11-1}.

Using sentiment analysis methods, we have found patterns in users' emotional behavior and 
observed the scale-free distribution of user activities in the whole forum and in singular threads as well as power law 
tails for the distribution of thread lengths and the number of unique users in a thread. 
At the level of the entire Forum, negative emotions boost users' activities, i.e. participants with more negative 
emotions write more posts. At the level of individual threads users that are more active in a specific thread 
tend to express there more negative emotions and seem to be the key agents  sustaining thread discussions. 
As a result, longer threads possess more negative emotional content. 

Internet slang calls people who frequently post inflammatory messages designed to increase strife among users 
\emph{trolls}. Our statistics linking increased individual activity with negative emotional attitudes
suggests that such users may be the source of \emph{flame wars}, in which many users resort to impolite language, easily detected by automated tools. Exchanges of angry posts between pairs of users raises 
the average emotional temperature of the debate and may encourage other users to adopt a similar tone, \textbf{creating}
a generally negative emotional content in a discussion thread. 
A comparison of the data and agent-based simulations suggest that much of the emotional bias comes from reactive messages,
especially prolonged quarrels between pairs of users with opposing views.

Papacharissi in \cite{papacharissi04-1} differentiated between impolite and uncivil messages. The first
group often uses vulgar language and would be easily detectable by sentiment analysis methods, while the second 
is defined as impeccably mannered but challenging the standards of opposing group, whole society or an individual.
Analyzing a small sample of 268 messages, Papacharissi could not provide statistical measures, but he describes specific 
cases when calm discussion could turn into a heated debate after one impolite or uncivil message. He mentioned examples 
of quarrels spurred by such postings, noting that they eventually calmed down. However, distinguishing the hot emotion of 
impolite posts from cold aggression of uncivil ones, he notes that authors of uncivil messages never apologized  nor 
retraced their comments. Moreover, the automated sentiment detection programs used by us could underestimate the amount 
of negative emotions, when expressed as politely worded sarcasm or provocations.

The simulation model that includes extended exchanges between pairs of users reproduces many of 
the observed characteristics of the BBC forum reasonably well and is offered as a possible explanation of the behaviour found. 
A similar model was 
used in \cite{sobkowicz10-1}, where thanks to combination of comment organization in the studied dataset and 
categorization analysis conducted by humans (as opposed to automated process) it has been possible to verify 
directly the `quarrel' model of user activity. In \cite{sobkowicz10-1}, for a highly controversial politics forum, 
over 70\% of messages were identified as disagreements, invectives and provocations, which may be compared to 
a large proportion of negative sentiments expressed in our data. Additionally, the politics forum was found to differ from other, 
less controversial ones (such as sport, science or computer self-help), in the fact that the more active a user was, the 
higher was the percentage of his/her comments posted within pairwise exchanges. 
The same characteristic was found in \cite{rafaeli97-1}, 
where most active users have also the highest percentage of reactive messages, reaching up to 87\%.

As quarrels are an important part of the simulated system, we undertook to check if they are 
found in the original data. The nature of our dataset did not allow automatic recognition of all quarrels. 
The time ordering process flattened the dataset, hiding this information. Full attribution would require 
a search through the text of the messages for tell-tales of pairwise exchanges of posts, for example 
references to other posts/users within a comment. A very simplified analysis based on the temporal proximity 
of the posts, looking for series of consecutive posts such as ABAB\ldots, where A and B are unique authors, 
gives an estimate of the quarrels at more than 40\%. A detailed analysis of a subset of threads shows that, 
especially for most active ones, the posts of pairs of agents are likely to be separated by other comments, 
so this statistic strongly underestimates the percentage of discussions.
This should be compared to simulations, where about 70\% of posts were parts of such pairwise discussions.
This is higher than the ratio of 52.5\% found by Rafaeli and Sudweeks \cite{rafaeli97-1}. The difference may be due 
to the use of randomly selected groups in \cite{rafaeli97-1}, which may have included some less controversial fora. 
This hypothesis is corroborated by a large percentage of information providing posts reported by Rafaeli and Sudweeks.

In discussions of emotions we must take into account the effects of anonymity. A review by Derks et. al. 
\cite{derks08-1} lists several studies which show that anonymous communication results in more 
uninhibited behavior and being more critical than in face to face conditions or when the author could be recognized. 
Interestingly, Kushin and Kitchener \cite{kushin09-1} have noted that participants in Facebook political 
discussions have shown high levels of negative emotions and uncivil behavior, even though much of the anonymity 
is removed by the available Facebook profiles. Some users actually expressed the belief that they are fully anonymous. 
On the other hand, when people know that they are not anonymous (such as e-mail communication or videoconferencing), 
they tend to be more restrained \cite{siegel86-1}. 
High levels of negative emotions have led Serfaty \cite{serfaty02-1} to question if controversial newsgroups 
really deserve to be classified as communities. If one defines a community as network of people built on closeness, 
shared identity, solidarity and common goals then this is indeed questionable. Of course, even the anonymous users form 
networked structures, with opponents often seeking out each other through various threads to continue the same 
quarrels again and again \cite{sobkowicz10-1}, but without a common goal the negative attitudes dominate. 

With these observations in mind, the way to moderate the emotional tone of news discussions could involve 
two measures: first, making the users internally aware of their traceability, even if they are anonymous towards 
each other; and second, to focus their attention on the topic of the news, rather than comments made by other users. 
This may be achieved by rearrangement of the user interface, making it easier to comment on source news and harder 
to comment on another user's post. Preliminary results of a study of the same political forum that was analyzed in 
\cite{sobkowicz10-1}, after such a change of user interface confirm this hypothesis 
[Sobkowicz, unpublished].

\section*{Acknowledgments}
The work was supported by EU FP7 ICT Project {\it Collective Emotions in Cyberspace - CYBEREMOTIONS}, European COST Action MP0801 {\it Physics of Competition and Conflicts} and Polish Ministry of Science Grant 1029/7.PR UE/2009/7 and Grant  578/N-COST/2009/0.  Plots 1, 2a, 2b  were first created by Jan Lorenz (Cyberemotions internal report). 
\section*{Appendix} 
We performed the analysis of sensitivity of numerical results on model parameters.
The parameters used in simulations interact in a rather complex way. It is possible, for 
example, to 'compensate' the increase in the number of neutral agents by increasing the 
probability of negative response to such agents and quarrel probability, within certain 
limits. 
The parameters final values were derived by looking first at  statistical properties of 
the BBC Forum, starting with the number of messages, distributions  of messages within a 
thread and from agiven user etc. In search for a good fit to all the observations, some 
of the parameters were modified only in rough way, for example the relative numbers of 
A/B/N agents and source messages. This was because we decided to use a single set of 
parameters for all discussions - a clear difference from reality, where in reality each 
discussion could gather different sets of users, with different ratios of view points. 
For this reason we used the verb "chosen" rather than "fitted" in our model description. 

An example of model sensitivity can be as follows.  Let us decrease the parameter $x_N$ 
from $0.91$ to $0.8$ and keep all other parameters of the model constant. It leads   to 
changes  of ratios  of comments with positive, neutral and negative emotional valencies 
from respectively $0.19$, $0.16$ and $0.65$ to $0.199$,$0.190$, $0.610$. A further decrease of $x_N$ 
to $0.5$ leads to corresponding values of $0.204$, $0.237$,$0.560$. On the other hand increasing 
the value of $x_N$ to 1 (all posts by neutrals provoke a negative emotion when commented 
by agents A and B) leads to ratios of  $0.201$, $0.144$, $0.655$ for emotionally positive, 
neutral and negative messages. 

However, in all cases the distribution of the emotion average as a function of thread 
length (as shown in Fig. \ref{ke}  is shifted up (for decreased values of $x_N$) or down (for 
increased $x_N$), but the distribution  does not loose its characteristic "peak" for $L$ 
values around $10$. 

Decreasing $p_r$ (keeping all other parameters constant) some how flattens the `peak' seen 
in Fig.\ref{ke}, but at the `cost' of decreasing the number of comments in the forum. For 
example using $p_r=0.85$ results in the 'peak' value of $\left<e\right>_L$ changing from $-0.18$ to$-0.21$, 
but the number of comments drops to 2 million, due to a smaller number of quarrels. A 
further decrease flattens the peak slightly more. At the same time, smaller $p_r$ values 
lead not only to smaller number of overall comment number (which may be compensated by 
increasing $p_c$) but also to wrong shapes of distribution of thread lengths and user 
activities. 

It is our belief that the difference between simulations and observations results from 
automatic assignment of emotions to posts based on the opinion of the authors. In 
reality, emotions are spurred often by specific wording of a comment, its incivility, 
abuse addressed ad personam etc. Let us consider a single highly negative remark 
directed to a source of news, that would not lead to a quarrel. Such a comment may set 
the 'tone' for the discussion, especially for short ones, when due to a small number of 
comments all users would see the offensive message and become irritated. To achieve the 
same levels of negative emotions, our agent based model relies on longer discussions and 
the role of quarrels in longer threads. This is in agreement with a strongly 
heavy-tailed distribution of thread lengths observed in the studied dataset (Fig \ref{2}A).


\bibliographystyle{model1-num-names}

\section*{Figure Legends}


\end{document}